\documentclass[%
superscriptaddress,
 amsmath,amssymb,
 aps,prb,
 twocolumn 
 ]{revtex4-2}

\usepackage{graphicx}
\usepackage{dcolumn}
\usepackage{bm}
\usepackage{xcolor}
\usepackage[caption=false]{subfig}
\usepackage{xfrac}
\usepackage{blindtext}
\usepackage{times}
\usepackage{comment}
\usepackage{simpler-wick}
\usepackage{hyperref}

\definecolor{darkblue}{HTML}{004D6B}
\definecolor{darkred}{HTML}{8c1515}
\hypersetup{
	colorlinks=true,
	urlcolor=darkblue,
	citecolor=darkred,
	linkcolor=darkblue,
	breaklinks
}

\begin{document}


\title{Perturbative Analysis of Quasi-periodic Patterning of Transmon Quantum Computers: Enhancement of Many-Body Localization}

\author{Evangelos Varvelis}
\affiliation{Institute for Quantum Information, RWTH Aachen University, 52056 Aachen, Germany}
\affiliation{J\"ulich-Aachen Research Alliance (JARA), Fundamentals of Future Information Technologies, 52425 J\"ulich, Germany}

\author{David P. DiVincenzo}
\affiliation{Institute for Quantum Information, RWTH Aachen University, 52056 Aachen, Germany}
\affiliation{J\"ulich-Aachen Research Alliance (JARA), Fundamentals of Future Information Technologies, 52425 J\"ulich, Germany}
\affiliation{Peter Gr\"unberg Institute, Theoretical Nanoelectronics, Forschungszentrum J\"ulich, 52425 J\"ulich, Germany}%


\date{\today}

\begin{abstract}
Recently it has been shown that transmon qubit architectures experience a transition between a many-body localized and a quantum chaotic phase. While it is crucial for quantum computation that the system remains in the localized regime, the most common way to achieve this has relied on disorder in Josephson junction parameters. Here we propose a quasi-periodic patterning of parameters as a substitute for random disorder. We demonstrate, using the Walsh-Hadamard diagnostic, that quasiperiodicity is more effective than disorder for achieving localization. In order to study the localizing properties of our new Hamiltonian for large, experimentally relevant system sizes, we use two complementary perturbation-theory schemes, one with respect to the many-body interactions and one with respect to hopping parameter of the free Hamiltonian.
\end{abstract}

\maketitle

\section{Introduction}

Despite immense advances in quantum computing using the superconducting qubit platform \cite{GoogleQuantumSupremacy,ChineseQuantumSupremacy,IBMQuantumCloud}, two-qubit gate fidelity remains a thorn in the side of further progress with these devices. One prominent source of these errors is quantum cross talk in the form of qubit $ZZ$ couplings \cite{ZZOriginal}, with $Z$ denoting the Pauli $z$ operator. This cross talk is the result of always-on coupling of qubits, even in idle mode. There are two primary strategies for dealing with these residual couplings, tunable coupling \cite{TunableCoupling} and static coupling between opposite anharmonicity qubits \cite{Ansari1,Ansari2}. Each of these comes with disadvantages: additional hardware overlay of couplers for the former and lower coherence time of capacitively shunted flux qubits for the latter. In real devices the existence of natural random disorder is unavoidable and most prominently present in the critical current of Josephson junctions. Even though in modern devices tuning of Josephson junctions is possible, even post fabrication, with laser annealing techniques \cite{LASIQ}, some degree of residual disorder remains. 

The many-body system formed by a network of $N$ Josephson qubits, with random disorder and fixed coupling, is a prime candidate for quantum chaos. Recently it has been established that there is in fact a phase transition between quantum chaotic and many body localization (MBL) for transmon arrays \cite{TransmonChaos} of this type. The phenomenology of this transition can be summarised by considering the diagonalized Hamiltonian of such a multiqubit system
\begin{equation}\label{TauHamiltonian}
H = \sum_{a\in\mathbb{B}^N}E_{a}\vert a\rangle\langle a\vert = \sum_{b\in\mathbb{B}^N}w_b Z_1^{b_1}\dots Z_N^{b_N},
\end{equation}
where $\mathbb{B}^N$ is the set of all bit-strings of length $N$, $w_b$ is a real coefficient corresponding to bit-string $b$, $b_i$ is the $i$-th digit of bit-string $b$, and $Z_i$ is the Pauli $z$ operator acting on the subspace of qubit $i$. The coefficients $w_b$ with a bit-string $b$ consisting of only two 1s in adjacent sites correspond, by definition, to the $ZZ$-couplings. Longer-range $ZZ$ couplings or higher-weight terms have generally been neglected (but see \cite{xu2024lattice}), a treatment that is consistent in the MBL phase, where we have an exponential hierarchy of these terms with respect to correlation range \cite{MBL1,MBL2}. This is in stark contrast with the chaotic regime however, where all of these terms are of the same order of magnitude. 

These systems can only be deep in the MBL phase due to the happenstance of disorder in the energies $E_J$ of the Josephson qubits.
In this paper we explore the stabilization of the MBL phase with a quasi-periodic potential replacing the randomly chosen disorder potential and determine that there is in fact a robust localized regime. We demonstrate this for a small system by obtaining the inverse participation ratio and the Walsh-Hadamard coefficients $w_b$ of Eq.~\eqref{TauHamiltonian} using exact diagonalization techniques. To obtain the latter we apply the Walsh-Hadamard transform on the spectrum of the system
\begin{equation}\label{WalshHadamard}
w_{b} = \frac{1}{2^N}\sum_{a\in\mathbb{B}^N}(-1)^{b\cdot\bar{a}}E_a,
\end{equation}
with $\bar{a}$ denoting a bit-string resulting from flipping each digit of bit-string $a$ \footnote{The reason for this bit flipping is the unfortunate convention in quantum information of having the ground state denoted by 1 and the first excited by 0, which we do not adopt here.}.

To extend our results to system sizes comparable to currently available devices, we will use a two-track analytic approach. On the one hand, we develop a bosonic variant of M\o{}ller-Plesset perturbation theory (MP) \cite{MollerPlessetOriginal}, treating the many-body interactions as the small parameter. On the other hand, we also use standard Rayleigh-Schr\"odinger perturbation theory (RS) in the hopping strength. For both schemes we obtain the energy levels of the qubit sector of the system, and consequently from these the Walsh-Hadamard coefficients from a direct application of the definition Eq.~\eqref{WalshHadamard}. The use of both schemes is necessitated by the fact that while MP perturbation theory scheme allows us to obtain longer-range correlations, it is only valid deep in the localized regime. On the contrary, RS perturbation theory scheme is more accurate for a broader region of the MBL regime, but obtaining longer-range correlations requires progressively higher orders of perturbation theory, rendering their calculation within this scheme impractical. 

\section{Metallic-Aubry-Andr\'e Model}

Here we focus on capacitively coupled transmon arrays. The minimal model Hamiltonian for such an array \cite{Transmons1,Transmons2} is
\begin{equation}\label{TransmonHamiltonian}
H = 4E_{\text{C}}\sum_{i=1}^{N}n_i^2 - \sum_{i=1}^{N}E_{\text{J}_i}\cos(\phi_i) + \lambda\sum_{\langle i,j\rangle}n_{i}n_{j},
\end{equation}
where $n_i$ is the Cooper-pair number of site $i$ and $\phi_i$ is the conjugate variable, corresponding to the superconducting phase. $E_{\text{C}}$ is the capacitive energy of each transmon, taken to be equal for all sites of the array. $E_{\text{J}_i}$ is the Josephson energy of site $i$, and $\lambda$ is the constant coupling strength between sites. We also assumed only nearest neighbor coupling.

After recasting the Hamiltonian of Eq.~\eqref{TransmonHamiltonian} in second quantization form and expanding the cosine term up to fourth order in order to include many-body interactions via the anharmonicity, we obtain the Bose-Hubbard approximation of our Hamiltonian,
\begin{equation}\label{BoseHubard}
H_{\text{BH}} = \sum_{i=1}^{N}\omega_i a_i^{\dagger}a_i + J\sum_{\langle i,j\rangle}a_{i}^{\dagger}a_{j} - \frac{E_{\text{C}}}{2}\sum_{i=1}^{N}a_i^{\dagger}a_i^{\dagger}a_ia_i.
\end{equation}
 We have also used a rotating wave approximation. Note that the ladder operators are bosonic --- we are not restricted to the single-excitation manifold of the Fock space. Strictly speaking, the new coupling strength $J$ would be bond-dependent and proportional to $\lambda\sqrt{\omega_i\omega_j}/E_{\text{C}}$ with $\omega_i = \sqrt{8E_{\text{J}_i}E_{\text{C}}} - E_{\text{C}}$. Here we have omitted this bond dependence in order to simplify the calculations. We find that including this dependence does not substantially alter our results.

\begin{figure}
    \centering
    \includegraphics[width = \columnwidth]{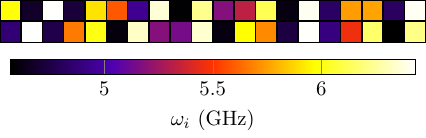}
    \caption{{\bf Metallic-Aubry-Andr\'e disorder potential:} Transmon frequency disorder potential $\omega(x_i,y_i)$ (Eq.~\eqref{MetalicRatioDisorder}) for a quasi-1D lattice of dimensions $2\times 20$. We use mean frequency $\langle\omega\rangle = 5.5\text{ GHz}$ and disorder strength $\Delta = 1\text{ GHz}$. The on-site frequencies are color-coded according to the given color bar and the values are given in GHz. The bottom row of the lattice corresponds to $y = 1$ and the top to $y = 2$. The leftmost sites of the lattice have $x = 1$ and the rightmost $x = 20$.}
    \label{Fig:DisorderPotential1}
\end{figure}

In order to design a frequency pattern $\omega_i$ for our transmon arrays, which should serve in place of a localizing disorder potential, the essential feature is to make it non repeating in order to avoid resonances. A secondary objective is to avoid having near-resonant sites in physical proximity for some specific lattice geometry. Our reasons for this will become clear later. Here we focus on a quasi-1D square lattice with dimensions $2\times L$, such as the one depicted in Fig.~\ref{Fig:DisorderPotential1}. Such a lattice geometry is already in use for actual quantum computing devices \cite{2xNLattice} and may also become even more relevant for future designs. 
Using integer-valued real space coordinates $(x_i,y_i)$ for site $i$, where $y$ is the short axis and $x$ the long axis of length $L>2$, we introduce the ``disorder" potential
\begin{equation}\label{MetalicRatioDisorder}
\omega_i = \langle\omega\rangle + \Delta\sqrt{2}\sin\left[\pi x_i\left(y_i + \sqrt{y_i^2 + 4}\right)\right].
\end{equation}
Here $\langle\omega\rangle$ is the central value around which the transmon frequencies vary with a strength that is determined by the sine function amplitude $\Delta$. More precisely the parameter definitions have been chosen such that in the thermodynamic limit ($N\rightarrow\infty$) we will have 
\begin{equation}
    \frac{1}{N}\sum_{i=1}^{N}\omega_i \rightarrow \langle\omega\rangle\quad\text{and}\quad\sqrt{\frac{1}{N}\sum_{i=1}^{N}(\omega_i-\langle\omega\rangle)^2} \rightarrow \Delta.
\end{equation}
In other words $\langle\omega\rangle$ is the average and $\Delta$ the standard deviation.

The inspiration behind using this potential is the Aubry-Andr\'e model commonly used in the study of 1D quasicrystals, where it exhibits a well studied transition between Anderson localized \cite{AndersonLocalizationOriginal,AndersonLocalizationReview} and delocalized phases \cite{AubryAndreOriginal,AubryAndre1D1,AubryAndre1D2,AubryAndre2D1,AubryAndre2D2}. As a matter of fact, treating the $y$ coordinate as a fixed parameter we recover the exact form of the Aubry-Andr\'e model. Simply put, along the $x$ direction, the disorder potential is Aubry-Andr\'e while changing the $y$ coordinate simply changes the period of the sine function. In order for the potential to be non-periodic, the periodicity of the sine function must be chosen so that it is incommensurate with the integer periodicity of the lattice. This is done here by making the periodicity irrational, and since we need the periodicity to be varying with $y$ we need a family of irrational numbers, hence our choice of the metallic ratios \cite{MetallicRatios}.

\begin{figure*}
    \centering
    \includegraphics[width = \textwidth]{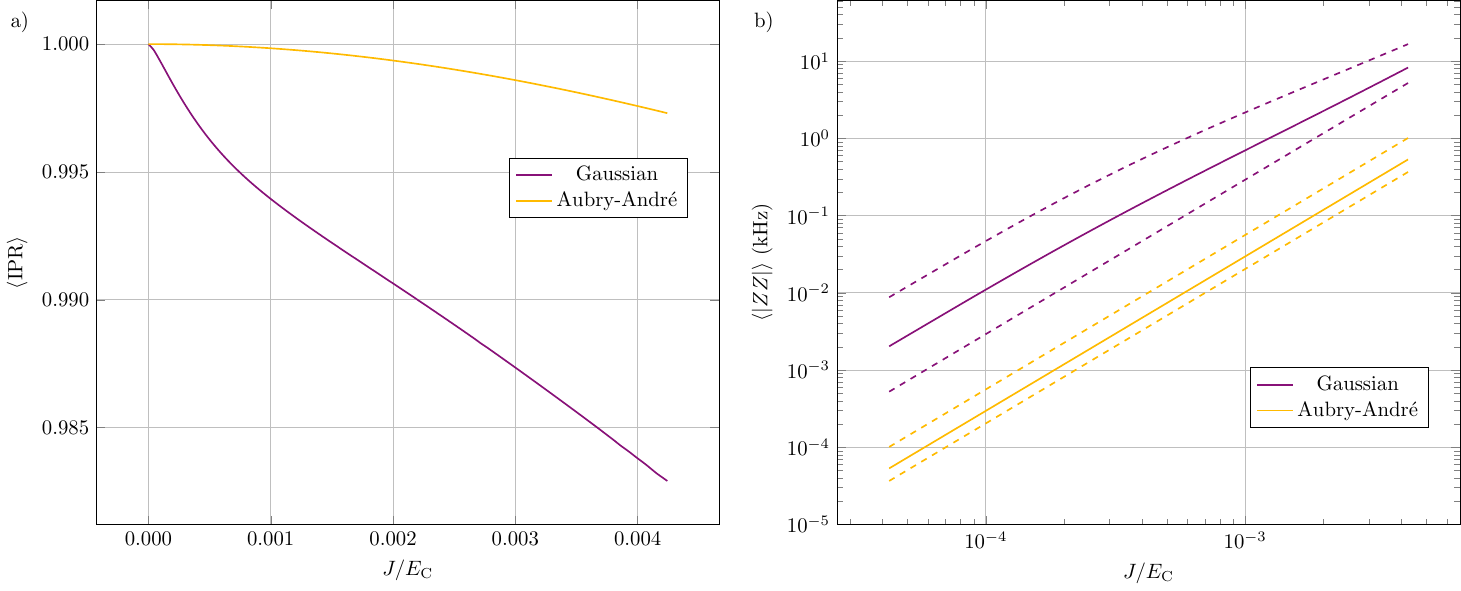}
    \caption{{\bf Random Gaussian disorder vs Metallic-Aubry-Andr\'e:} Exact diagonalization results for averaged (a) Inverse participation ratio (IPR) and (b) Walsh-Hadamard coefficients as a function of the hopping strength $J$ in units of the anharmonicity $E_\text{C}$ for MAA and random Gaussian disorder with matching standard deviation. The averaging is done over (a) all eigenstates and (b) Walsh-Hadamard coefficients of weight 2 and correlation range $\ell = 1$ (nearest-neighbor $ZZ$ coefficients). Particullarly for the $ZZ$ coefficients plot, the dashed lines indicate the range of the $ZZ$ coefficients of the solid line with the same color. For the case of the Gaussian random disorder there is additional averaging over 100 different realisations. The parameter values used here are $E_{\text{C}}=0.33\text{ GHz}$, $\langle\omega\rangle = 5.5\text{ GHz}$, $\Delta = 1\text{ GHz}$ and the coupling $J$ is varied over the range $0$ to $1.5\text{ MHz}$. The system size for both cases is $2\times 3$.}
    \label{Fig:GaussianDisorder}
\end{figure*}

With the Metallic-Aubry-Andr\'e (MAA) model as our choice of  the disorder potential, our first goal is to establish how well it performs in localizing our system compared to random disorder.  We have first performed this comparison for a small system of size $2\times 3$, which is manageable with exact diagonalization. The results are reported in Fig.~\ref{Fig:GaussianDisorder} and confirm that our model outperforms random disorder of the same strength.

\subsection{Perturbation in the anharmonicity}

Using Hamiltonian Eq.~\eqref{BoseHubard} as an effective description of our system, we will obtain the Walsh-Hadamard coefficients perturbatively in the anharmonicity. All the results that will be presented in this section are derived in Appendix \ref{appendixMP}. The first two terms of the Hamiltonian in Eq.~\eqref{BoseHubard} describe the \textit{non-interacting} part of the Hamiltonian. We call it non-interacting in the sense that since it is quadratic in the ladder operators, in the eigenbasis it should obtain the form of uncoupled harmonic oscillators. In the \textit{bare basis}, the non-interacting part of the Hamiltonian is however not diagonal, and in order to use perturbation theory we first need to transform the Hamiltonian to the \textit{dressed basis} defined by the relation
\begin{equation}\label{SingleParticleCreationOperators}
c_{\mu}^{\dagger}\vert 0\rangle \stackrel{!}{=} \left(\sum_{i=1}^{N}\psi_{\mu}(x_i,y_i)a_i^{\dagger}\right)\vert 0\rangle,
\end{equation}
where $c_{\mu}^{\dagger}$ creates a single-excitation eigenstate $\vert\psi_{\mu}\rangle$ of the non-interacting Hamiltonian. For clarity, we reserve Latin indices for the bare basis and Greek indices for the dressed basis. 

Since we are only interested in the localised regime of the system, we could perform the transformation of Eq.~\eqref{SingleParticleCreationOperators} perturbatively as well (in the coupling $J$). However, this adds one additional layer of complexity to our final expressions, without leading to any particular new insights. Therefore we choose to obtain the single-particle sector spectrum numerically and use these results as input to our derived analytic expressions from second order perturbation theory in the anharmonicity $E_{\text{C}}$.

In the dressed basis, our Hamiltonian is recast into the form
\begin{equation}\label{DressedHamiltonian}
H_{\text{DBH}} = \sum_{\mu=1}^{N}\varepsilon_{\mu}c_{\mu}^{\dagger}c_{\mu} - \frac{E_\text{C}}{2}\sum_{\alpha,\beta,\mu,\nu =1}^{N}V_{\alpha\beta\mu\nu} c_\alpha^{\dagger} c_\beta^\dagger c_\mu c_\nu,
\end{equation}
where $\varepsilon_{\mu}$ is the energy of a single excitation on site $\mu$ of the non-interacting Hamiltonian and we have also defined the 4-point correlation tensor 
\begin{equation}
    V_{\alpha\beta\mu\nu} = \sum_{i=1}^{N}\psi_\alpha(x_i,y_i)\psi_\beta(x_i,y_i)\psi_\mu(x_i,y_i)\psi_\nu(x_i,y_i),
\end{equation}
and we have made explicit use of the fact that our eigenstates are real. For sufficiently weak transmon coupling $J$, the single-excitation eigenstates should be exponentially localised around lattice site $\mu$ with coordinates $(x_{\mu},y_{\mu})$ and the dressed basis should be nearly identical to the bare basis $c_{\mu}^{\dagger} \approx a_{\mu}^{\dagger}$ and by extension for the energy levels as well: $\varepsilon_\mu\approx \omega_\mu$. Therefore, even though the capacitive energy $E_{\text{C}}$ is not the smallest energy scale in our system, the use of perturbation theory is justified as long as we are in the transmon regime since $E_{\text{C}}/\varepsilon_\mu \approx E_{\text{C}}/\omega_\mu \sim\sqrt{E_{\text{C}}/E_{\text{J}_\mu}}\ll 1$. 

\begin{figure*}
    \centering
    \includegraphics[width = \textwidth]{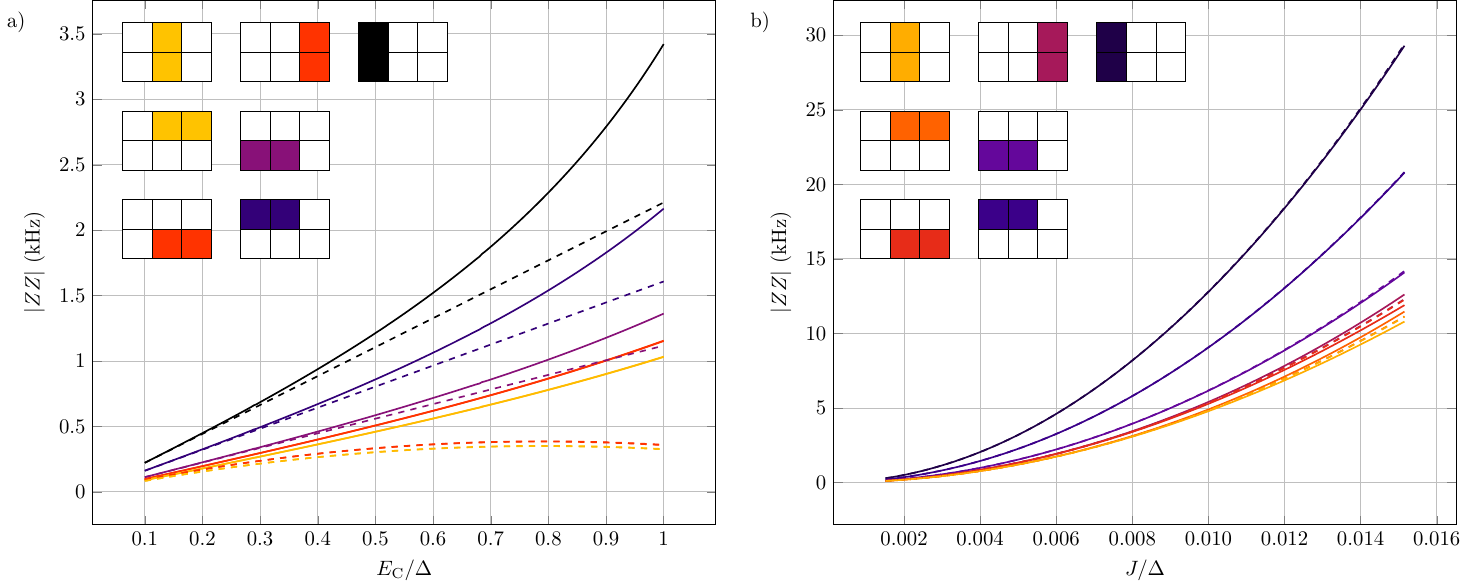}
    \caption{{\bf Exact diagonalization vs.~perturbation theory for the Metallic-Aubry-Andr\'e model:} Plots of all the nearest-neighbor $ZZ$ coefficients as a function of the small parameter $E_\text{C}$ in (a) and $J$ in (b), expressed in units of disorder strength $\Delta$. Exact diagonalization results are represented by solid lines and perturbation theory with dashed lines. The parameter values used for panel (a) are $\Delta = 1\text{ GHz}$, $J = 2.5\text{ MHz}$ and the anhramonicity $E_\text{C}$ is varied over the range $0.1\Delta$ to $\Delta$. In panel (b) are $E_\text{C} = 0.33\text{ GHz}$, $\Delta = 2E_\text{C}$  and the hopping strength $J$ is varied over the range $1$ to $10\text{ MHz}$. For both cases $\langle\omega\rangle = 5.5\text{ GHz}$ and the system size is $2\times 3$. Coefficients are color coded with respect to the corresponding site pair, see legend.}
    \label{Fig:EDvsPT}
\end{figure*}

For the Walsh-Hadamard coefficients of Eq.~(\ref{WalshHadamard}), we only need to obtain the perturbed energy levels that correspond to qubit states
\begin{equation}\label{DressedStates}
\vert b\rangle = \prod_{\mu=1}^{N}(c_{\mu}^{\dagger})^{b_{\mu}}\vert 0\rangle,
\end{equation}
where $b_{\mu}$ is the $\mu$-th binary digit of bit string $b$. By allowing the bit-string digits to be undetermined parameters of possible value 0 or 1, we can obtain the energy correction for an arbitrary qubit state of any $N$-transmon array. The only thing that we need in order to calculate the involved amplitudes in the perturbation theory are the generalised Wick contraction rules for operators with binary undetermined exponents given by
\begin{equation}\label{WickRules}
\wick{\c1{c}_{\mu}^{b_{\alpha}}(\c1{c}_{\nu}^{\dagger})^{b_{\beta}}} = \delta_{\mu\nu}\delta_{b_{\alpha},b_{\beta}},
\end{equation}
while all other possible contractions are vanishing. It is straightforward to convince oneself of the validity of this relation simply by considering all 4 possible combinations of values for $b_\alpha$ and $b_\beta$. With these rules in hand, the relevant amplitudes can be calculated inductively using the method described in \cite{WickContractions}. See Appendices~\ref{appendixMP1} and \ref{appendixMP2} for an analytic derivation of these results.

Due to the linearity of the Walsh-Hadamard transformation, we can apply it separately at each order of perturbation theory for the qubit-sector energy levels, yielding a series expansion for the Walsh-Hadamard coefficients themselves:
\begin{equation}
    w_b = w_b^{(0)} + w_b^{(1)} + w_b^{(2)}  + \dots
\end{equation}
Furthermore, the analytical expressions for the energy levels contain terms that are proportional to products of a few bit-digits, particularly up to 3-digit products for the second order in perturbation theory that we calculated here. Such $m$-digit products transform like
\begin{subequations}
\begin{align}
    \Pi_{\mu_1\mu_2\dots\mu_m}(b) &= \frac{1}{2^N}\sum_{a\in\mathbb{B}^N}(-1)^{b\cdot\bar{a}}a_{\mu_1}a_{\mu_2}\dots a_{\mu_m}\\
    &= \frac{1}{2^m}\prod_{i\neq\mu_1,\mu_2,\dots,\mu_m}\bar{b}_i\label{WalshHadamardProduct}.
\end{align}
\end{subequations}
From Eq.~\eqref{WalshHadamardProduct}, it becomes apparent that, if the Walsh-Hadamard coefficient corresponds to a bit string $b$ of weight more than $m$, then $\Pi_{\mu_1\dots\mu_{m}}(b) = 0$.
As a direct consequence of this, first-order perturbation theory can only yield corrections for Walsh-Hadamard coefficients up to weight 2 and second order perturbation theory up to weight 3 (see Appendix \ref{appendixMP3} for more details). The exact form of these coefficients is 
\begin{subequations}
\begin{align}
w_{b}^{(0)} &= \sum_{\mu}\varepsilon_{\mu}\Pi_{\mu}(b),\label{ZerothOrderResU}\\
w_{b}^{(1)} &= \sum_{\mu<\nu}\mathcal{E}_{\mu\nu}\Pi_{\mu\nu}(b),\label{FirstOrderResU}\\
w_{b}^{(2)} &= \sum_{\stackrel{\mu<\nu}{\alpha,\beta}}\mathcal{D}_{\mu\nu\alpha\beta}\Pi_{\mu\nu}(b) + \sum_{\stackrel{\mu<\nu<\alpha}{\beta}}\mathcal{S}_{\mu\nu\alpha\beta}\Pi_{\mu\nu\alpha}(b),\label{SecondOrderResU}
\end{align}
\end{subequations}
where we have used the tensor definitions
\begin{subequations}
\begin{align}
\mathcal{E}_{\mu\nu} &= 2E_\text{C}V_{\mu\mu\nu\nu},\label{ETensor}\\
\mathcal{D}_{\mu\nu\alpha\beta} &= \frac{E_\text{C}^2}{2}\frac{\vert V_{\mu\nu\alpha\beta}\vert^2}{\varepsilon_\mu + \varepsilon_\nu - \varepsilon_\alpha - \varepsilon_\beta},\label{DTensor}\\
\mathcal{J}_{\mu\nu\alpha\beta} &= E_\text{C}^2\frac{V_{\mu\mu\alpha\beta}V_{\nu\nu\alpha\beta}}{\varepsilon_\alpha - \varepsilon_\beta},\label{JTensor}\\
\mathcal{S}_{\mu\nu\alpha\beta} &= 4!(\mathcal{D}_{(\mu\nu\alpha)\beta} + \mathcal{J}_{(\mu\nu\alpha)\beta}).\label{STensor}
\end{align}
\end{subequations}
The instances which are according to these definitions divergent, $\mathcal{D}_{\mu\nu\mu\nu}, \mathcal{D}_{\mu\nu\nu\mu}$ and $\mathcal{J}_{\mu\nu\alpha\alpha}$, do not appear in the expressions for $w_b$. We have also used the notation for the fully symmetric component of a tensor $T$ with respect to the indices in parentheses, which for our case simplifies due to the explicit symmetry of our tensors $\mathcal{D}$ and $\mathcal{J}$ with respect to the first two indices to
\begin{equation}\label{Symmetrization}
    T_{(\mu\nu\alpha)\beta} = \frac{1}{3}(T_{\mu\nu\alpha\beta}+T_{\mu\alpha\nu\beta}+T_{\nu\alpha\mu\beta}).
\end{equation}

By using the MAA scheme, we set up a quasi-random disorder potential without resonances and with well separated near-resonant sites. While the four-term denominator of $\mathcal{D}$ creates some dangers for perturbation theory, its numerators have a counteracting effect. They are proportional to the 4-point function $V_{\mu\nu\alpha\beta}$ of the single-particle eigenvectors involving the same states as the ones in the denominator. Anderson localization theory ensures that these correlations decay exponentially with range.

With this perturbation theory, we can move on to obtain Walsh-Hadamard coefficients for a much larger system of dimensions $2\times 20$. Before doing that, however, we examine the accuracy of our perturbation theory by comparing it with exact diagonalization results. For this comparison, still restricted to the $2\times 3$ system size, see Fig.~\ref{Fig:EDvsPT}. It is evident that the agreement of the two results is restricted to a rather small parameter range. Even though the second order perturbation theory is very accurate for the energy levels, with an error of $\sim 10^{-1}\text{ kHz}$ for eigenenergies spanning a few tens of GHz, the error is of the same order of magnitude as the Walsh-Hadamard coefficients of weight 2, and is about 2 orders of magnitude larger that coefficients of weight 3. This is why we only report the weight 2 coefficients here. Unfortunately the accuracy of the energy levels is not found to be improved by introducing higher-order terms \cite{MollerPlessetAccuracy}; our perturbation theory is equivalent to that of the $\varphi^4$ theory, which is known to have a vanishing radius of convergence. Already at third order of perturbation theory, the disagreement with the exact diagonalization results starts to increase. 

\begin{figure}
    \centering
    \includegraphics[width = \columnwidth]{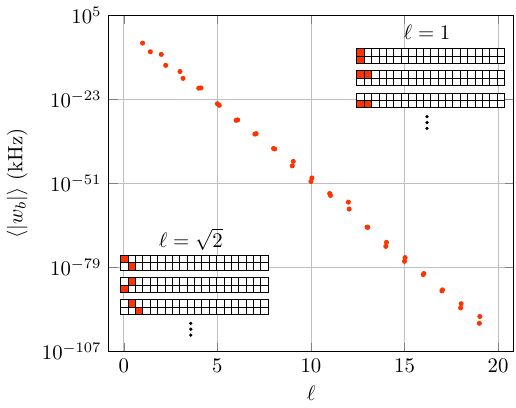}
    \caption{{\bf Exponential hierarchy of the Walsh-Hadamard coefficients for the Metallic-Aubry-Andr\'e model:} Plot of averaged Walsh-Hadamard coefficients of weight 2 as a function of the correlation range $\ell$ for a $2\times 20$ quasi-1D lattice. Averaging is done among Walsh-Hadamard coefficients of the same correlation range. The parameter values used here are $J = 1\text{ MHz}$, $\langle\omega\rangle = 5.5\text{ GHz}$, $E_{\text{C}} = 0.33\text{ GHz}$ and $\Delta = 3E_\text{C}$. The inset is a visual representation of the relevant Walsh-Hadamard coefficient bit-strings for correlation range $\ell = 1$ and $\ell = \sqrt{2}$ on the lattice. Filled boxes represent excited sites while empty ones ground state}
    \label{Fig:WH2x20}
\end{figure}

Despite these difficulties, we obtain meaningful results for the Walsh-Hadamard coefficients and the correct order of magnitude within the parameter range $\Delta\gtrsim 4E_\text{C}$, as can be seen in Fig.~\ref{Fig:EDvsPT}. Therefore we proceed to obtain the weight-2 coefficients using perturbation theory for the much larger $2\times 20$ system, beyond the size that is easily accessible to exact diagonalization. The results for this calculation are reported in Fig.~\ref{Fig:WH2x20}. They confirm the expectation that these Walsh-Hadamard coefficients exhibit a strong hierarchy of values, decreasing exponentially with range; this is as expected within many-body localization theory (see \cite{TransmonChaos}).

\subsection{Perturbation in the hopping strength}

We have seen that treating the many-body interactions perturbatively yields information for the many-body localization of the system but only for extremely weak anharmonicity or strong disorder. In order to overcome this restricted radius of convergence we need to employ a different perturbative scheme without the pathologies of the $\varphi^4$ theory. Here we will treat perturbatively instead the hopping strength $J$ of the system Hamiltonian of Eq.~\eqref{BoseHubard}. 

For this scheme the unperturbed Hamiltonian is diagonal in the bare basis with qubit states represented by
\begin{equation}\label{BareStates}
\vert b\rangle = \prod_{\mu=1}^{N}(a_{\mu}^{\dagger})^{b_{\mu}}\vert 0\rangle,
\end{equation}
and therefore we can apply immediately RS perturbation theory without going to the dressed basis. We again perform a second-order perturbation theory calculation for the qubit-sector energy levels (see Appendix~\ref{appendixRS} for derivation) and apply the Walsh-Hadamard transformation on the analytic expressions for the energy levels in order to derive the following results:
\begin{subequations}
\begin{align}
w_{b}^{(0)} &= \sum_{j}\omega_{j}\Pi_{j}(b),\label{ZerothOrderResJ}\\
w_{b}^{(1)} &= 0,\label{FirstOrderResJ}\\
w_{b}^{(2)} &= J^2\sum_{\langle i<j \rangle}\left[\frac{4E_\text{C} }{E_\text{C}^2 - \delta_{ij}^2} + \frac{2(b_i-b_j)}{\delta_{ij}}\right]\Pi_{ij}(b),\label{SecondOrderResJ}
\end{align}
\end{subequations}
where we have defined the detuning between sites $i$ and $j$ as $\delta_{ij} = \omega_i-\omega_j$. With the bracketed summation indices $\langle i < j\rangle$ we denote summation with respect to the indices $i$ and $j$ of nearest neighbour sites only and with a fixed ordering to avoid double counting pairs. The bit-digit products $\Pi_j(b)$ and $\Pi_{ij}(b)$ are the ones already defined in Eq.~\eqref{WalshHadamardProduct}.

In panel b of Fig.~\ref{Fig:EDvsPT} we present a comparison of the perturbation theory results of Eqs.~\eqref{ZerothOrderResJ}-\eqref{SecondOrderResJ} with the numerical values of the Walsh-Hadamard coefficients obtained with exact diagonalization for a small system of dimensions $2\times 3$. The RS perturbation scheme is strikingly accurate especially when compared to the accuracy of the MP result in panel a of Fig.~\ref{Fig:EDvsPT}. Even more remarkably, the RS perturbation theory scheme manages to capture the features of our model significantly further away from the strongly localized regime. This is evident from the fact that within the presented range of $J$ values the $ZZ$ coefficients have managed to climb to an order of magnitude higher values than the corresponding case for MP perturbation theory, with virtually no drop in accuracy.

This drastic improvement in accuracy with our new perturbation theory scheme however comes at the cost of a reduced capability of probing the exponential hierarchy of the Walsh-Hadamard coefficients. Note that up to second order in perturbation theory we can only obtain corrections for bit-strings of up to weight 2 using the same argument given right after the definition of the bit-digit products Eq.~\eqref{WalshHadamardProduct}. Additionally, the summation in Eq.~\eqref{SecondOrderResJ} is over nearest-neighbour pairs only. From the last two observations we can infer that only $ZZ$ coefficients can be estimated with second order perturbation theory. Even for weight 2 bit-strings of longer correlation range $\ell >1$, we would need to go to higher orders of perturbation theory.

But another advantage of the RS perturbation scheme is the significantly lower computational cost for obtaining the $ZZ$ coefficients. According to Eq.~\eqref{SecondOrderResJ}, calculating a particular $ZZ$ coefficient to second order, one needs to only calculate a single term for $i,j$, referring to the sites of the $ZZ$ coefficient under consideration. In contrast, using the MP perturbation scheme for the second-order $ZZ$ coefficients one needs to invoke a summation of $\sim N^2$ terms. As a result of that we are able to calculate the second order $ZZ$ coefficients for a system size of $2\times 2000$ in a matter of seconds using the RS scheme, while MP requires several minutes to yield the same results for a $2\times20$ system like the one presented in Fig.~\ref{Fig:WH2x20}. 

\section{Conclusion}

We have demonstrated that it is possible to localize a many-body quantum computing system without the use of random disorder but rather with a deterministically designed, nonperiodic potential. We believe that the disorder potential we studied here is not yet optimal and that meticulous frequency pattern engineering should play a crucial role in the design of future quantum computing architectures. Our new perturbation-theory scheme can be used as a guide for the properties a frequency pattern should possess or avoid. Despite the limited accuracy of the MP perturbation scheme, we have demonstrated that it is possible to obtain useful analytical results for the Walsh-Hadamard coefficients of large many-body systems. We believe that accuracy can ultimately be improved with a renormalized perturbation theory in the anharmonicity. Finally, we need to stress that the results for the $2\times 20$ lattice are beyond the realm of what is attainable with exact diagonalization techniques, and of course the $2\times 2000$ lattice even far more so. The main impediment for increasing the size further is the exponential scaling of the number of Walsh-Hadamard coefficients themselves, which is $2^N$. However, if we instead restrict the calculation to only low-weight coefficients with small correlation range, then we can obtain reliable results using the RS scheme for extremely large systems, without any compromise in accuracy, as long as we remain in the desired localized regime.

\begin{acknowledgements}
We acknowledge support from the Deutsche Forschungsgemeinschaft (DFG) 
under Germany's Excellence Strategy Cluster of Excellence Matter and Light for Quantum Computing (ML4Q) EXC 2004/1 390534769.    
\end{acknowledgements}

\appendix

\section{M\o{}ller-Plesset Perturbation Theory}\label{appendixMP}

We start with a Hamiltonian of the form given in Eq.~\eqref{DressedHamiltonian} but without making any assumption for the form of the tensor $V_{\alpha\beta\mu\nu}$, besides that it is real. However there are two symmetries that will simplify our calculation significantly without loss of generality. The first one is the fact that
\begin{align}
\sum_{\alpha,\beta,\mu,\nu}V_{\alpha\beta\mu\nu}c_{\alpha}^{\dagger}c_{\beta}^{\dagger}c_{\mu}c_{\nu} &= \sum_{\alpha,\beta,\mu,\nu}V_{\alpha\beta\mu\nu}c_{\beta}^{\dagger}c_{\alpha}^{\dagger}c_{\mu}c_{\nu}\nonumber\\ 
&= \sum_{\alpha,\beta,\mu,\nu}V_{\beta\alpha\mu\nu}c_{\alpha}^{\dagger}c_{\beta}^{\dagger}c_{\mu}c_{\nu},
\end{align}
where in the first step we used the commutativity of the bosonic operators and in the second step we performed a summation index relabeling. Even though this property does not guarantee that $V_{\alpha\beta\mu\nu} = V_{\beta\alpha\mu\nu}$, it does suggest that if $V_{\alpha\beta\mu\nu}$ has an antisymmetric component with respect to the first two indices then that component yields no contribution to the sum. Since we can always decompose $V_{\alpha\beta\mu\nu}$ to a symmetric and an antisymetric component it means we can assume without loss of generality that $V_{\alpha\beta\mu\nu}$ can always be symmetrized with respect to the first two indices without changing the Hamiltonian of the system. The exact same argument can be made for the last two indices. The second argument follows along similar lines 
\begin{align}
    \sum_{\alpha,\beta,\mu,\nu}V_{\alpha\beta\mu\nu}c_{\alpha}^{\dagger}c_{\beta}^{\dagger}c_{\mu}c_{\nu} &= \sum_{\alpha,\beta,\mu,\nu}V_{\alpha\beta\mu\nu}c_{\nu}^{\dagger}c_{\mu}^{\dagger}c_{\beta}c_{\alpha} \nonumber\\
    &= \sum_{\alpha,\beta,\mu,\nu}V_{\mu\nu\alpha\beta}c_{\alpha}^{\dagger}c_{\beta}^{\dagger}c_{\mu}c_{\nu},
\end{align}
where in the first step we used the Hermiticity of the sum and in the second step we performed a summation index relabeling. Once again, this does not imply $V_{\alpha\beta\mu\nu} = V_{\mu\nu\alpha\beta}$, however any component antihermitian with respect to the index pair exchange would have vanishing contribution to the sum and hence $V_{\alpha\beta\mu\nu}$ can be thought as symmetric with respect to the index pair exchange without loss of generality. In summary our symmetries for the tensor are
\begin{equation}\label{InteractionSymmetries}
    V_{\alpha\beta\mu\nu} = V_{\beta\alpha\mu\nu} = V_{\alpha\beta\nu\mu} = V_{\mu\nu\alpha\beta}.
\end{equation}
The tensor for the system we study in the main text has more symmetries than these and therefore what we present here is a more generic case.

Since the ladder operators are bosonic the system has an infinite number of eigenstates. However, for the Walsh-Hadamard coefficients, we are only interested in the finite subspace of qubit states of Eq.~\eqref{DressedStates}. For this family of states the unperturbed energy levels take the form
\begin{equation}\label{ZerothOrderCorrectionRes}
    E_{b}^{(0)} = \langle b\vert H_0\vert b\rangle = \sum_\mu E_{\mu}b_{\mu}
\end{equation}
and starting with this we can now obtain the energy levels of the generalized Hamiltonian with the usual Rayleigh-Schr\"odinger perturbation theory.

\subsection{First Order Energy Correction}\label{appendixMP1}

At first order in perturbation theory we will have
\begin{equation}\label{FirstOrderDef}
    E_b^{(1)} = \sum_{\alpha,\beta,\mu,\nu}V_{\alpha\beta\mu\nu}\langle b\vert c_{\alpha}^{\dagger}c_{\beta}^{\dagger}c_{\mu}c_{\nu}\vert b\rangle.
\end{equation}
If we substitute the state of Eq.~\eqref{DressedStates} in this definition, the calculation of the first order correction boils down to a single vacuum expectation value, which in turn can be evaluated by means of Wick contractions according to the rules presented in Eq.~\eqref{WickRules}. In order to do this we need to fully contract the operators. We will do this using the permanent method described in \cite{WickContractions}, which consists of calculating the permanent of the matrix whose elements corresponding to all possible Wick contractions between an annihilation operator, corresponding to a row, and a creation operator, corresponding to a column. Using the basis $\lbrace c_{\mu},c_{\nu},c_{1}^{b_1},\dots,c_{N}^{b_N}\rbrace$ for the rows and $\lbrace c_{\alpha}^{\dagger},c_{\beta}^{\dagger},(c_{1}^{\dagger})^{b_1},\dots,(c_{N}^{\dagger})^{b_N}\rbrace$ for the columns we can write Eq.~\eqref{WickRules} in matrix form
\begin{equation}
    \mathcal{C}_{\alpha\beta\mu\nu}(b) = 
    \begin{pmatrix}
        0 & 0 & b_1\delta_{\mu,1} & \dots & b_N\delta_{\mu,N} \\
        0 & 0 & b_1\delta_{\nu,1} & \dots & b_N\delta_{\nu,N} \\
        b_1\delta_{\alpha,1} & b_1\delta_{\beta,1} & 1 & \dots & 0 \\
        \vdots & \vdots & \vdots & \ddots & \vdots \\
        b_N\delta_{\alpha,N} & b_N\delta_{\beta,N} & 0 & \dots & 1
    \end{pmatrix}.
\end{equation}

For the first order correction we will have from Eq.~\eqref{FirstOrderDef}
\begin{equation}
    E_b^{(1)} = \sum_{\alpha,\beta,\mu,\nu}V_{\alpha\beta\mu\nu}~\text{perm}\left(\mathcal{C}_{\alpha\beta\mu\nu}(b)\right).
\end{equation}
By definition, the permanent of a $d\times d$ matrix is the sum over all possible products consisted of $d$ elements of the matrix with no two of them sharing a row or column. From the first two rows, only elements past column 2 have a non vanishing contribution to the product therefore, for example, if we select elements $2+\rho$ and $2+\sigma$ from the first and second row respectively, we will have a factor of $b_{\rho}b_{\sigma}\delta_{\mu\rho}\delta_{\nu\sigma}$ with the restriction that $\rho\neq\sigma$ which can be enforced by multiplying this expression with $\vert\varepsilon_{\rho\sigma}\vert$ with $\varepsilon$ denoting the fully antisymetric tensor. Having selected an element from columns $2+\rho$ and $2+\sigma$ means that we have effectively removed these entire columns and therefore at the corresponding rows the only possible choices that lead to a non vanishing contribution are from columns 1 and 2 yielding either $b_{\rho}b_{\sigma}\delta_{\alpha\rho}\delta_{\beta\sigma}$ or $b_{\rho}b_{\sigma}\delta_{\alpha\sigma}\delta_{\beta\rho}$ . With the first two rows and columns of the matrix removed we can only choose elements from the identity matrix at the bottom right of the contraction matrix. The permanent for our two-body interaction case yields
\begin{equation}
    E_b^{(1)} = \sum_{\stackrel{\alpha,\beta,\mu,\nu}{\rho,\sigma}}V_{\alpha\beta\mu\nu}b_{\rho}b_{\sigma}\vert\varepsilon_{\rho\sigma}\vert\delta_{\mu\rho}\delta_{\nu\sigma}(\delta_{\alpha\rho}\delta_{\beta\sigma} + \delta_{\alpha\sigma}\delta_{\beta\rho}).
\end{equation}
Performing the summation with respect to $\rho,\sigma,\alpha$ and $\beta$ and using the symmetries of the interaction potential Eq.~\eqref{InteractionSymmetries} we obtain
\begin{equation}\label{FirstOrderCorrectionRes}
    E_b^{(1)} = \sum_{\mu<\nu}\mathcal{E}_{\mu\nu}b_{\mu}b_{\nu}.
\end{equation}
where we used the tensor definition
\begin{equation}\label{ETensorDef}
    \mathcal{E}_{\mu\nu} = 4V_{\mu\nu\mu\nu}.
\end{equation}
Note that according to the symmetries of $V_{\alpha\beta\mu\nu}$ given in Eq.~\eqref{InteractionSymmetries} the tensor $\mathcal{E}$ should be symmetric
\begin{equation}
    \mathcal{E}_{\mu\nu} = \mathcal{E}_{\nu\mu}
\end{equation}
Finally performing the Walsh-Hadamard transformation on Eq.~\eqref{FirstOrderCorrectionRes} we obtain the result presented in the main text Eq.~\eqref{FirstOrderResU}. For more details on this see Appendix~\ref{appendixMP3}.

\subsection{Second Order Energy Correction}\label{appendixMP2}
We now proceed with the second order correction given by
\begin{align}
    E_b^{(2)} = & V_{\alpha_1\beta_1\mu_1\nu_1}V_{\alpha_2\beta_2\mu_2\nu_2}\times\nonumber\\
     & \times\sum_{m\neq b}^{}\frac{\langle b\vert c_{\alpha_1}^{\dagger}c_{\beta_1}^{\dagger}c_{\mu_1}c_{\nu_1}\vert m\rangle\langle m\vert c_{\alpha_2}^{\dagger}c_{\beta_2}^{\dagger}c_{\mu_2}c_{\nu_2}\vert b\rangle}{E_b^{(0)}-E_m^{(0)}},\label{SecondOrderDef}
\end{align}
where we have used Einstein summation convention here to lighten the notation. We have a new complication here, since the eigenstates $\vert m\rangle$ of the free Hamiltonian are not necessarily qubit states anymore and as a result Wick contraction rules for arbitrary exponents become significantly more complicated. However we can circumvent this using the following argument. The state resulting from the application of $c_{\alpha_2}^{\dagger}c_{\beta_2}^{\dagger}c_{\mu_2}c_{\nu_2}$ on the qubit state $\vert b\rangle$ is still an eigenstate $\vert n_2\rangle$ of the free Hamiltonian although it might not be normalised anymore
\begin{equation}    c_{\alpha_2}^{\dagger}c_{\beta_2}^{\dagger}c_{\mu_2}c_{\nu_2}\vert b \rangle = \lambda_2 \vert n_2\rangle.
\end{equation}
A similar argument follows for the bracket
\begin{equation}
    \langle b\vert c_{\alpha_1}^{\dagger}c_{\beta_1}^{\dagger}c_{\mu_1}c_{\nu_1} = \lambda_1 \langle n_1\vert.
\end{equation}
and therefore we can write using Eq.~\eqref{SecondOrderDef}
\begin{subequations}
\begin{align}
    E_b^{(2)} &= \lambda_1\lambda_2 V_{\alpha_1\beta_1\mu_1\nu_1}V_{\alpha_2\beta_2\mu_2\nu_2}\sum_{m\neq b}\frac{\langle n_1\vert m\rangle\langle m\vert n_2\rangle}{E_b^{(0)}-E_m^{(0)}}\\
    &= \lambda_1\lambda_2 V_{\alpha_1\beta_1\mu_1\nu_1}V_{\alpha_2\beta_2\mu_2\nu_2}\sum_{m\neq b}\frac{\delta_{n_1,m}\delta_{n_2,m}}{E_b^{(0)}-E_m^{(0)}}\\
    &= \lambda_1\lambda_2 V_{\alpha_1\beta_1\mu_1\nu_1}V_{\alpha_2\beta_2\mu_2\nu_2}\left. \frac{\delta_{n_1,n_2}}{E_b^{(0)}-E_{n_2}^{(0)}}\right\vert_{n_{2}\neq b}\\
    &= V_{\alpha_1\beta_1\mu_1\nu_1}V_{\alpha_2\beta_2\mu_2\nu_2}\left. \frac{\lambda_1\lambda_2\langle n_1\vert n_2\rangle}{E_b^{(0)}-E_{n_2}^{(0)}}\right\vert_{n_{2}\neq b}.
\end{align}
\end{subequations}
Finally since by definition the state $\vert n_2 \rangle$ is obtained from $\vert b\rangle$ by extracting two excitations from sites $\mu_2$ and $\nu_2$ and adding two at sites $\alpha_2$ and $\beta_2$ we will have 
\begin{equation}
    E_{n_2} = E_{b} - \varepsilon_{\mu_2} - \varepsilon_{\nu_2} + \varepsilon_{\alpha_2} + \varepsilon_{\beta_2}
\end{equation}
and in total
\begin{align}
    E_b^{(2)} = & V_{\alpha_1\beta_1\mu_1\nu_1}V_{\alpha_2\beta_2\mu_2\nu_2}\times\nonumber\\
    & \times \left. \frac{\langle b\vert c_{\alpha_1}^{\dagger}c_{\beta_1}^{\dagger}c_{\mu_1}c_{\nu_1}c_{\alpha_2}^{\dagger}c_{\beta_2}^{\dagger}c_{\mu_2}c_{\nu_2}\vert b\rangle}{\varepsilon_{\mu_2} + \varepsilon_{\nu_2} - \varepsilon_{\alpha_2} - \varepsilon_{\beta_2}}\right\vert_{\lbrace \alpha_2,\beta_2\rbrace\neq\lbrace \mu_2,\nu_2\rbrace},\label{SecondOrderCorrection}
\end{align}
where $\lbrace \alpha_2,\beta_2\rbrace\neq\lbrace \mu_2,\nu_2\rbrace$ is meant in the sense of set inequality. Once again we have an amplitude involving only a single qubit state and therefore we can use the Wick rules from the first order calculation leading directly to the contraction matrix
\begin{equation} \small
    \begin{pmatrix}
        0 & 0 & \delta_{\mu_1\alpha_2} & \delta_{\mu_1\beta_2} & b_1\delta_{\mu_1,1} & \dots & b_N\delta_{\mu_1,N} \\
        0 & 0 & \delta_{\nu_1\alpha_2} & \delta_{\nu_1\beta_2} & b_1\delta_{\nu_1,1} & \dots & b_N\delta_{\nu_1,N} \\
        0 & 0 & 0 & 0 & b_1\delta_{\mu_2,1} & \dots & b_N\delta_{\mu_2,N} \\
        0 & 0 & 0 & 0 & b_1\delta_{\nu_2,1} & \dots & b_N\delta_{\nu_2,N} \\
        b_1\delta_{\alpha_1,1} & b_1\delta_{\beta_1,1} & b_1\delta_{\alpha_2,1} & b_1\delta_{\beta_2,1} & 1 & \dots & 0 \\
        \vdots & \vdots & \vdots & \vdots & \vdots & \ddots & \vdots \\
        b_N\delta_{\alpha_1,N} & b_N\delta_{\beta_1,N} & b_N\delta_{\alpha_2,N} & b_N\delta_{\beta_2,N} & 0 & \dots & 1
    \end{pmatrix}.
\end{equation}
For this case, since the effective four-body interaction is not normal ordered already, we have additional contractions. There are multiple ways to deal with this, namely normal ordering the interaction term using the bosonic algebra, the method of reduced permanents or considering cases as we did before for the first order correction. The results are summarised in the following expression
\begin{equation}\label{SecondOrderCorrectionRes}
    E_{b}^{(2)} = \sum_{\stackrel{\alpha<\beta}{\mu,\nu}}\mathcal{D}_{\alpha\beta\mu\nu}b_{\alpha}b_{\beta} + \sum_{\stackrel{\alpha<\beta<\mu}{\nu}}\mathcal{S}_{\alpha\beta\mu\nu}b_{\alpha}b_{\beta}b_{\mu},
\end{equation}
where we have also introduced the tensor definitions
\begin{subequations}
\begin{align}
    \mathcal{D}_{\alpha\beta\mu\nu} &= 
    \begin{cases} 
        \displaystyle{\frac{2\vert V_{\alpha\beta\mu\nu}\vert^2}{\varepsilon_{\alpha} + \varepsilon_{\beta} - \varepsilon_{\mu} - \varepsilon_{\nu}},} & \lbrace\alpha,\beta\rbrace\neq\lbrace\mu,\nu\rbrace\\ 
        \displaystyle{0,} & \lbrace\alpha,\beta\rbrace = \lbrace\mu,\nu\rbrace .
    \end{cases}\label{DTensorDef}\\
    \mathcal{J}_{\alpha\beta\mu\nu} &= 
    \begin{cases}       
        \displaystyle{\frac{4V_{\alpha\mu\alpha\nu}V_{\beta\mu\beta\nu}}{\varepsilon_{\mu} - \varepsilon_{\nu}},} & \mu \neq \nu \\
        \displaystyle{0,} & \mu = \nu
    \end{cases},\label{JTensorDef}\\
    \mathcal{S}_{\alpha\beta\mu\nu} &= 4!(\mathcal{D}_{(\alpha\beta\mu)\nu} + \mathcal{J}_{(\alpha\beta\mu)\nu}).\label{STensorDef}    
\end{align}
\end{subequations}
with the indices in parenthesis denoting the symmetrization of Eq.~\eqref{Symmetrization} in the main text. In conjunction with Eq.~\eqref{InteractionSymmetries} we can conclude the following symmetries
\begin{equation}\label{DSymetries}
    \mathcal{D}_{\alpha\beta\mu\nu} = \mathcal{D}_{\beta\alpha\mu\nu} = \mathcal{D}_{\alpha\beta\nu\mu} = -\mathcal{D}_{\mu\nu\alpha\beta},
\end{equation}
\begin{equation}\label{JSymetries}
    \mathcal{J}_{\alpha\beta\mu\nu} = \mathcal{J}_{\beta\alpha\mu\nu} = -\mathcal{J}_{\alpha\beta\nu\mu},
\end{equation}
and $\mathcal{S}_{\mu\nu\alpha\beta}$ is symmetric under any permutation of the first three indices.

The antisymmetric property of the tensors $\mathcal{D}$ and $\mathcal{J}$ is the reason why in the second order correction of Eq.~\eqref{SecondOrderCorrectionRes} we have no four-bit terms. Indeed, Eq.~\eqref{SecondOrderCorrectionRes} is the result of the summation over all possible full contractions of the amplitude in Eq.~\eqref{SecondOrderCorrection}. For the final step, we once again apply the Walsh-Hadamard transformation on Eq.~\eqref{SecondOrderCorrectionRes} to obtain the result of the main text presented in Eq.~\eqref{SecondOrderResU}. We will give more details on this in the following section.

\subsection{Second Order Walsh-Hadamard Coefficients}\label{appendixMP3}

With the Rayleigh-Schr\"{o}dinger perturbation theory results of Eqs.~\eqref{FirstOrderCorrectionRes} and \eqref{SecondOrderCorrectionRes} for the qubit state energy levels we are now able to derive the Walsh-Hadamard coefficients via direct application of the definition in Eq.~\eqref{WalshHadamard}. Furthermore, since the transformation is linear we can perform it term by term
\begin{equation}
    w_{b}^{(n)} = \frac{1}{2^N}\sum_{q\in\mathbb{B}^N} (-1)^{b\cdot\overline{q}}E_q^{(n)}.
\end{equation}
We start with the zeroth order term of Eq.~\eqref{ZerothOrderCorrectionRes}
\begin{equation}
    w_{b}^{(0)} = \frac{1}{2^N}\sum_{\mu=1}^{N}\sum_{q\in\mathbb{B}^N} (-1)^{b\cdot\overline{q}}E_{\mu}q_{\mu}.
\end{equation}
In order to carry out the Boolean summation we split the bit-string summation over a product of bit-digit summations and use the following property of functions of binary variables
\begin{equation}\label{BinaryFExpansion}
    f(b) = \overline{b}f(0) + bf(1),
\end{equation}
from which it follows that
\begin{equation}\label{BinSumsP1}
    \sum_{q = 0}^{1}(-1)^{b\overline{q}} = 2\overline{b}\ ,\ \sum_{q = 0}^{1}(-1)^{b\overline{q}}q = 1,
\end{equation}
\begin{equation}\label{BinSumsP2}
    \text{and } \sum_{q = 0}^{1}(-1)^{b\overline{q}}\overline{q} = 1 - 2b
\end{equation}
As an example we present the summation explicitly for zeroth order case
\begin{subequations}
\begin{align}
    w_{b}^{(0)} &= \sum_{\mu=1}^{N}\frac{E_{\mu}}{2^N}\sum_{q\in\mathbb{B}^N} (-1)^{b\cdot\overline{q}}q_{\mu}\\
    &= \sum_{\mu=1}^{N}\frac{E_{\mu}}{2^N}\left(\prod_{\stackrel{\rho=1}{\rho\neq\mu}}^{N}\sum_{q_\rho=0}^{1} (-1)^{b_{\rho}\cdot\overline{q}_{\rho}}\right)\left(\sum_{q_{\mu}=0}^{1} (-1)^{b_{\mu}\cdot\overline{q}_{\mu}}q_{\mu}\right)\\
     &= \sum_{\mu=1}^{N}\frac{E_{\mu}}{2}\prod_{\stackrel{\rho=1}{\rho\neq\mu}}^{N}\overline{b}_{\rho} = \sum_{\mu=1}^{N}E_{\mu}\Pi_\mu(b),\label{WalshHadamardZero}
\end{align}
\end{subequations}
where in the last step we used the definition of Eq.~\eqref{WalshHadamardProduct}. For the first and second order terms using the same procedure we can derive the results presented in Eq.~\eqref{FirstOrderResU} and Eq.~\eqref{SecondOrderResU}.

The products of the flipped bit digits pose a sharp cutoff for the weight of Walsh-Hadamard coefficients we can estimate at a finite order of perturbation theory. Namely for a bit-string $b^{(m)}$ of weight $m$, the product of all flipped bit digits excluding $k$, with $m>k$,  will be vanishing for any set of excluded flipped digits since at least one of them will be zero. Therefore, to second order in perturbation theory, we can only obtain corrections for the Walsh-Hadamard coefficients of weight up to 3. For the non vanishing cases with $m\leq k$, assuming that the non zero digits of the bit-string are located at positions $\ell_1$ through $\ell_m$ in ascending order, the product will yield one if $\lbrace \ell_1,\dots,\ell_m\rbrace$ is a subset of the excluded digits. The above can be summarised in the following expression
\begin{equation}\label{FlippedDigitProduct}
    \Pi_{\mu_1\dots\mu_k}(b^{(m)}) = \frac{\theta(k-m)}{2^k}\sum_{s\in [\mu_1,\dots,\mu_k]^m}\prod_{j = 1}^{m}\delta_{\ell_j,s(j)},
\end{equation}
with $\theta$ denoting the Heaviside step function with the convention $\theta(0) = 1$ and $[\mu_1,\dots,\mu_k]^m$ denotes the set of all oriented subsets of length $m$ of the set $\lbrace\mu_1,\dots,\mu_n\rbrace$. As an example the flipped digit product for a bit-string of weight $m=2$ with $k=3$ excluded digits yields
\begin{align}
    \Pi_{\mu_1\mu_2\mu_3}(b^{(2)}) &= \frac{1}{8}(\delta_{\ell_1}^{\mu_1}\delta_{\ell_2}^{\mu_2} + \delta_{\ell_1}^{\mu_2}\delta_{\ell_2}^{\mu_1} + \delta_{\ell_1}^{\mu_1}\delta_{\ell_2}^{\mu_3} + \nonumber\\ 
    &\qquad + \delta_{\ell_1}^{\mu_3}\delta_{\ell_2}^{\mu_1} + \delta_{\ell_1}^{\mu_2}\delta_{\ell_2}^{\mu_3} + \delta_{\ell_1}^{\mu_3}\delta_{\ell_2}^{\mu_2}).
\end{align}
We used upper indices here only for presentation purposes and no additional context.

\section{Derivation of RS perturbation theory results in the hopping strength}\label{appendixRS}

Starting from the Hamiltonian in Eq.~\eqref{BoseHubard} and working in the bare basis the qubit energy levels to zeroth order in the hopping strength $J$ will be
\begin{equation}
    E_{b}^{(0)} = \sum_{i=1}^{N}\omega_i \langle b\vert a_i^{\dagger}a_i\vert b\rangle = \sum_{i=1}^{N}\omega_i b_i.
\end{equation}
Note that contributions from the anharmonicity term are trivially vanishing for qubit states. First order corrections are also trivially vanishing
\begin{equation}
    E_{b}^{(1)} = J\sum_{\langle ij\rangle} \langle b\vert a_i^{\dagger}a_j\vert b\rangle = 0.
\end{equation}

At second order in perturbation theory we will have the correction
\begin{equation}\label{HoppingSecondOrder}
E_b^{(2)} = J^2\sum_{m\neq b}\sum_{\stackrel{\langle ij\rangle}{\langle k\ell\rangle}}\frac{\langle b\vert a_k^{\dagger}a_\ell\vert m\rangle\langle m\vert a_i^{\dagger}a_j\vert b\rangle}{E_b^{(0)}-E_m^{(0)}}    
\end{equation}
We note here that the $\vert m\rangle$ states are generic bosonic Fock states and therefore their unperturbed energy is going to include anharmonic effects
\begin{align}
    E_{m}^{(0)} &= \sum_{i=1}^{N}\omega_i \langle m\vert a_i^{\dagger}a_i\vert m\rangle - \frac{E_\text{C}}{2}\sum_{i=1}^{N}\langle m\vert a_i^{\dagger}a_i^{\dagger}a_ia_i\vert m\rangle\nonumber\\
    &= \sum_{i=1}^{N}\omega_i m_i - \frac{E_\text{C}}{2}\sum_{i=1}^{N}m_i(m_i-1).
\end{align}

The numerator of the second order correction in Eq.~\eqref{HoppingSecondOrder} can be calculated using a similar argument as the one we used for the second order correction of the MP perturbation scheme in Appendix \ref{appendixMP2}. However it is more instructive and useful for calculating higher order corrections to give a diagrammatic interpretation for them as quasiparticles hopping on the lattice.

More precisely one can interpret the product $\langle b\vert a_k^{\dagger}a_\ell\vert m\rangle\langle m\vert a_i^{\dagger}a_j\vert b\rangle$ as the situation where we start with a configuration of quasi-particles corresponding to the state $\vert b\rangle$. Since $\vert b\rangle$ is a qubit state there is at most one quasiparticle at each side. The first hop occurs by annihilating a quasiparticle at site $j$ and creating it again at an adjacent site $i$. This first hop brings us to the intermediate state $\vert m\rangle$ which may or may not be a qubit state. Next we finish our walk by hopping one quasiparticle from site $\ell$ to site $k$ which should bring us back to the configuration of state $\vert b\rangle$. Therefore the only possible paths are quasiparticles hopping to adjacent sites and then return to the initial positions
\begin{equation}
E_b^{(2)} = J^2\sum_{m\neq b}\sum_{\langle ij\rangle}\frac{\langle b\vert a_j^{\dagger}a_i\vert m\rangle\langle m\vert a_i^{\dagger}a_j\vert b\rangle}{\sum_{k=1}^{N}\omega_k(b_k-m_k) + \frac{E_\text{C}}{2}m_k(m_k-1)},
\end{equation}
where we have also substituted the unperturbed energy levels.

In order to calculate the weight of each step in this 2-step path we consider all cases. For the first step we initially annihilate a quasiparticle at site $j$ and since the starting state is a qubit state we will get a factor of $\sqrt{b_j} = b_j$. This last equality is derived from the property of Eq.~\eqref{BinaryFExpansion}. We will use this property repeatedly throughout this derivation. Next we need to create a quasiparticle at the adjacent site $i$. Since $\vert b\rangle$ is a qubit state it is initially at most singly occupied, in which case we get a factor of $\sqrt{2}$ other wise we get a factor of $1$. We can summarize this in the factor $2^{b_i/2}$. Note that with the arrival of the quasiparticle at site $j$ the state $\vert m\rangle$ is restricted to have $m_k = b_k$ for all $k$ except from $m_j = 0$ and $m_i = b_i+1$. For the second step we start from site $i$ which is now either singly or doubly occupied. With a similar argument as the one we used for the arrival of the quasiparticle at site $i$ we obtain a factor of $2^{b_i/2}$ for its departure and since its initial position is certainly empty the last factor is $1$. In total
\begin{equation}
    E_b^{(2)} = J^2\sum_{\langle ij\rangle}\frac{b_j 2^{b_i}}{\omega_jb_j - \omega_i + \frac{E_\text{C}}{2}b_i(b_i+1)}.
\end{equation}
Using the property Eq.~\eqref{BinaryFExpansion} we can simplify this into
\begin{equation}
    E_b^{(2)} = J^2\sum_{\langle ij\rangle}\left(\frac{\overline{b_i} b_j}{\omega_j-\omega_i} + \frac{2b_i b_j}{\omega_j - \omega_i +E_\text{C}}\right).
\end{equation}
As a final step since every pair of sites $i,j$ appears twice in the sum we fix an arbitrary order for the summations indices and have
\begin{align}
    E_b^{(2)} &= J^2\sum_{\langle i<j\rangle}\left(\frac{\overline{b_j} b_i - \overline{b_i} b_j}{\omega_i-\omega_j} + \frac{4b_i b_jE_\text{C}}{E_\text{C}^2 - (\omega_i-\omega_j)^2}\right).
\end{align}

Applying to the expressions that we derived here for the energy corrections the Walsh-Hadamard transformation Eq.~\eqref{WalshHadamard} and using the properties Eqs.~\eqref{BinSumsP1} and \eqref{BinSumsP2} we can obtain the Eqs.~\eqref{ZerothOrderResJ}-\eqref{SecondOrderResJ} of the main text. 

The next non-vanishing contribution is at fourth order of perturbation theory. However, in the quasiparticle walks picture that we presented for the second order calculation, there are now much more ways for the particle to move around which means we need to consider many different cases, each one contributing a rather complicated summation, therefore presenting this result here does not offer any particular new insights and for computational purposes we deem the numerical calculation of the higher order corrections advantageous although analytical results are obtainable. 

\bibliography{references.bib}

\end{document}